\documentclass[american,aps,showpacs,showkeys,twocolumn]{revtex4}
\usepackage{amsmath}
\usepackage{graphicx}

\makeatletter
\@ifundefined{showcaptionsetup}{}{%
 \PassOptionsToPackage{caption=false}{subfig}}
\usepackage{subfig}
\makeatother

\usepackage{babel}
\begin{document}

\title{Analytic solution of the fractional advection diffusion equation
for the time-of-flight experiment in a finite geometry}

\author{B.W. Philippa}
\email{bronson.philippa@my.jcu.edu.au}

\affiliation{School of Engineering and Physical Sciences, James Cook University,
Townsville 4811, Australia}

\author{R.D. White}

\affiliation{School of Engineering and Physical Sciences, James Cook University,
Townsville 4811, Australia}

\author{R.E. Robson}

\affiliation{School of Engineering and Physical Sciences, James Cook University,
Townsville 4811, Australia}

\keywords{fractional diffusion; anomalous diffusion; time of flight experiment;
organic semiconductors }

\pacs{05.40.Fb, 73.50.-h, 05.60.-k}
\begin{abstract}
A general analytic solution to the fractional advection diffusion
equation is obtained in plane parallel geometry. The result is an
infinite series of spatial Fourier modes which decay according to
the Mittag-Leffler function, which is cast into a simple closed form
expression in Laplace space using the Poisson summation theorem. An
analytic expression for the current measured in a time-of-flight experiment
is derived, and the sum of the slopes of the two respective time regimes
on logarithmic axes is demonstrated to be $-2$, in agreement with
the well known result for a continuous time random walk model. The
sensitivity of current and particle number density to variation of
experimentally controlled parameters is investigated in general, and
the results applied to analyze selected experimental data.
\end{abstract}
\maketitle

\section{Introduction}

Modern solid state electronics is largely based upon inorganic, crystalline
materials, such as silicon and germanium, the transport properties
of which are generally well understood \cite{Kittel1996SolidState}.
The same applies to gaseous electronics, for which there is a one-to-one
correspondence with crystalline condensed matter \cite{Robson1985}.
On the other hand, organic semiconductors are attracting increasing
interesting because of their desirable properties, such as transparency,
flexibility, and the prospect of economic advantage over inorganic
electronics \cite{Forrest2004}. Organic materials, which may be amorphous,
exhibit electrical properties which are generally qualitatively and
quantitatively quite different from inorganic materials \cite{Scher1991}.
For example, charge carriers in a time-of-flight experiment exhibit
long lived, spatially dispersed structures. Furthermore, the roles
of the mobility and diffusion coefficients, $\mu$ and $D_{L}$ respectively,
are not at all clear cut, as they are in crystalline structures or
gases. Such anomalous or {}``dispersive'' behavior arises because
the scattering of charge carriers may be accompanied by trapping in
localized states for times $\tau$, as determined by a {}``relaxation
function'' $\phi(\tau)$, which has an asymptotic time dependence
$\sim\tau^{-\gamma}$, with fractional exponent $\gamma$. 

The recent interest in {}``fractional kinetics'' derives mainly
from the seminal paper of Scher and Montroll \cite{Scher1975}, whose
discussion in terms of a continuous time random walk has spawned an
extensive literature in its own right \cite{Scher1991,Compte1996,Compte1998,Metzler2000,Barkai2001,Barkai2002,Metzler2004}.
In this literature, it is often assumed that the charge carrier number
density $n(z,t)$ may be found as the solution of a fractional diffusion
equation, which for present purposes we will refer to as the {}``Caputo''
form of the fractional advection diffusion equation:
\begin{equation}
_{0}^{C}D_{t}^{\gamma}n+W\frac{\partial n}{\partial z}-D_{L}\frac{\partial^{2}n}{\partial z^{2}}=0,\label{eq:caputo-fade}
\end{equation}
where $_{0}^{C}D_{t}^{\gamma}$ is the Caputo fractional partial derivative
with respect to $t$ of order $\gamma$. The Caputo derivative (see
Appendix A) accounts for trapping in localized states. This is appropriate
for a thin sample of amorphous material confined between two large
plane parallel boundaries, with all spatial variation confined to
the normal direction, which defines the $z$ axis of a system of coordinates.
In addition it is assumed that the small signal limit prevails, and
that both the drift velocity $W=\mu E$ , also directed along the
$z$ axis, and the longitudinal diffusion coefficient $D_{L}$ derive
entirely from an externally applied field $E$. For non-dispersive
transport, $\gamma\to1$ such that $_{0}^{C}D_{t}^{\gamma}n\rightarrow\frac{\partial n}{\partial t}$,
and Eq. \eqref{eq:caputo-fade} assumes the familiar classical form
\cite{Huxley1974}. The present article focuses on new techniques
for solution of Eq. \eqref{eq:caputo-fade} for the purposes of better
understanding the factors influencing experiment. 

Before proceeding with the detailed analysis, it is important to bear
in mind that Eq. \eqref{eq:caputo-fade} is only approximate. Just
as the kinetic theory of classical charge carrier transport in crystalline
semiconductors and gases has been developed to a sophisticated level
through solution of Boltzmann's kinetic equation, a more general and
accurate picture of anomalous transport in amorphous media should
be obtained through solution of a fractional kinetic equation in phase
space, in which the microscopic collision operator accounts for scattering
and trapping processes. Projection onto configuration space is achieved
by integration over velocity space, yielding (with approximations)
Eq. \eqref{eq:caputo-fade} plus expressions for macroscopic properties
such as $\mu$ and $D_{L}$. The phase space approach is beyond the
scope of the present work, and the reader is referred to \cite{Robson2005}
for such considerations.

Whatever the medium, gaseous or condensed matter, crystalline or amorphous,
the advection diffusion equation \eqref{eq:caputo-fade} is usually
assumed to provide the link between theory and experiment, its limitations
not withstanding. Thus, on the one hand, solution of the Boltzmann
kinetic equation provides theoretical values of $\mu$ and $D_{L}$,
and on the other, solution of Eq. \eqref{eq:caputo-fade} for $n(z,t)$,
with appropriate boundary and initial conditions, enables experimental
data to be unfolded to furnish empirical values of the same transport
properties. Comparison of theoretically derived and experimentally
measured transport properties then gives information about the fundamental
microscopic nature of the interaction of charge carriers with the
medium including the trapping/detrapping process. This procedure is
standard for electrons and ion {}``swarms'' in gases \cite{Huxley1974},
but application of the idea to amorphous media awaits the further
development of fractional Boltzmann phase space kinetics. That is
part of our long term theoretical program, but in the meantime, we
focus in the present article on the more practical imperative of developing
an accurate and efficient means of solving Eq. \eqref{eq:caputo-fade}.

To this end, a simple and numerically efficient solution of Eq. \eqref{eq:caputo-fade}
would be highly desirable. Previously reported solutions of fractional
diffusive systems for bounded media have been expressed in terms of
infinite series solutions \cite{Agrawal2002,Metzler2000b,Metzler2000}.
We show that the series solution to Eq. \eqref{eq:caputo-fade} with
absorbing boundaries may be collapsed into a simple closed form solution
in Laplace space by building upon the experience gained in solution
of the non-dispersive diffusion equation in gaseous electronics, specifically,
for the pulsed radiolysis drift tube experiment \cite{Robson1985}.
The structure of this article is as follows: In Section II, we model
the time of flight experiment \cite{Tiwari2009} and obtain a formal
analytic solution of Eq. \eqref{eq:caputo-fade} as a series of Mittag-Leffler
functions, which is cast into a tractable form, suitable for practical
purposes, using the Poisson summation theorem. In Section III, we
express the current measured in a time-of-flight experiment in terms
of this analytic solution, and show analytically that sums of the
slopes in distinct time regimes add up to -2 on a log-log plot, as
first predicted by Scher and Montroll \cite{Scher1975} and as observed
in many experiments \cite{Scher1991}. In Section IV, we explore the
way that current varies with experimental parameters, and go on to
fit selected experimental data. We show that our solution demonstrates
the power-law decay characteristic of dispersive transport.

\section{Analytic solutions of the fractional diffusion equation}

In this article, we will use Eq. \eqref{eq:caputo-fade} to model
a disordered semiconductor in a time of flight experiment \cite{Tiwari2009}.
The relationship between the various forms of the fractional advection
diffusion equation using both Caputo and Riemann-Liouville forms of
the fractional derivative operator are discussed in Appendix A. A
one dimensional equation, such as \eqref{eq:caputo-fade}, is appropriate
for a thin sample of disordered material confined between two large
plane parallel boundaries, which we shall take to be at $z=0$ and
$L$ respectively. All spatial variation is confined to the normal
direction, which defines the $z$ axis of a system of coordinates.
In addition it is assumed that the small signal limit prevails, and
that both the drift velocity $W=\mu E$ (where $\mu$ is the mobility)
and the longitudinal diffusion coefficient $D_{L}$ derive entirely
from an externally applied field $E$.

In the idealized time-of-flight experiment, a sharp pulse of $n_{0}$
charge carriers is released from a source plane $z=z_{0}$ at time
$t=t_{0},$ i.e., 
\begin{equation}
n(z,t_{0})=n_{0}\delta(z-z_{0}),\label{INITIAL}
\end{equation}
and the fractional advection diffusion equation is solved using the
methods and techniques described below. The solution for other experimental
arrangements, e.g., for sources distributed in space and/or emitting
for finite times, can be found by appropriate integration of this
fundamental solution over $z_{0}$ and/or $t_{0}$ respectively. The
solution for perfectly absorbing boundaries, for which
\begin{equation}
n(0,t)=0=n(L,t)\label{BCs}
\end{equation}
is 
\begin{equation}
n(z,t)=n_{0}\sum_{m=1}^{\infty}\varphi_{m}(z)E_{\gamma}\left(-\omega_{m}\left(t-t_{0}\right)^{\gamma}\right),\label{eq:SolnA}
\end{equation}
where the spatial modes are
\[
\varphi_{m}(z)\equiv\frac{e^{\lambda\left(z-z_{0}\right)}}{L}\left(\cos\left[k_{m}(z-z_{0})\right]-\cos\left[k_{m}(z+z_{0})\right]\right),
\]
and where\begin{subequations} 
\begin{align}
\lambda & \equiv\frac{W}{2D_{L}}\\
\omega_{m} & \equiv D_{L}\left(\lambda^{2}+k_{m}^{2}\right)\\
k_{m} & \equiv\frac{m\pi}{L}.
\end{align}
\end{subequations}In Eq. \eqref{eq:SolnA}, $E_{\gamma}(z)$ is the
Mittag-Leffler function of order $\gamma$: 
\begin{eqnarray}
E_{\alpha,\beta}(z) & \equiv & \sum_{k=0}^{\infty}\frac{z^{k}}{\Gamma(\alpha k+\beta)}\label{eq:mlf-power-series}\\
E_{\alpha}(z) & \equiv & E_{\alpha,1}(z).\nonumber 
\end{eqnarray}

Equation \eqref{eq:SolnA} gives an exact solution, however, this
expression is somewhat difficult to manipulate due to the presence
of the Mittag-Leffler function. Furthermore, a large number of terms
are needed for this series to converge, and the numerical evaluation
of the Mittag-Leffler to suitable precision is computationally difficult.

As is well known, fractional models obey a correspondence principle,
where non-fractional behavior is recovered in appropriate limits.
In this case, in the limit $\gamma\to1$ the Mittag-Leffler function
reduces to an exponential, i.e. $E_{1}(z)=e^{z}$, and \eqref{eq:SolnA}
reduces to Eq. (3b) in Ref. \cite{Robson1985}. In the classical,
non-fractional limit \cite{Robson1985}, it was shown that the series
convergence could be substantially improved through application of
the Poisson summation theorem (PST): 
\begin{equation}
\sum_{m=-\infty}^{\infty}f(mT)=\frac{1}{T}\sum_{m=-\infty}^{\infty}F\left(\frac{m}{T}\right),
\end{equation}
where $F(k)$ is the Fourier transform of $f(x)$. This article will
demonstrate that the PST can also be applied to the fractional advection
diffusion equation with similar benefits. Attempting to apply the
PST directly to Eq. \eqref{eq:SolnA} results in an intractable Fourier
transform involving the Mittag-Leffler function. On the other hand,
the Mittag-Leffler function has a simple Laplace domain representation.
Transformed into Laplace space, Eq. \eqref{eq:SolnA} becomes
\begin{equation}
\bar{n}(z,s)=n_{0}\sum_{m=1}^{\infty}\varphi_{m}(z)\frac{s^{\gamma-1}}{s^{\gamma}+\omega_{m}},\label{eq:SolnB}
\end{equation}
where without loss of generality we have taken $t_{0}=0$.

Applying the Poisson summation theorem to Eq. \eqref{eq:SolnB} gives
the equivalent form 
\begin{multline}
\bar{n}(z,s)=\alpha e^{\lambda z}\sum_{m=-\infty}^{\infty}\bigg[e^{-\beta\left|2Lm-\left(z-z_{0}\right)\right|}\\
  -e^{-\beta\left|2Lm-\left(z+z_{0}\right)\right|}\biggr],\label{eq:SolnC}
\end{multline}
where the space-independent parameters $\alpha$ and $\beta$ are
defined as 
\begin{align}
\alpha(s) & \equiv\frac{n_{0}s^{\gamma-1}e^{-\lambda z_{0}}}{2\sqrt{D_{L}}\sqrt{s^{\gamma}+D_{L}\lambda^{2}}}\\
\beta(s) & \equiv\frac{\sqrt{s^{\gamma}+D_{L}\lambda^{2}}}{\sqrt{D_{L}}}.\label{eq:beta}
\end{align}

Simplifying Eq. \eqref{eq:SolnC}, we obtain the closed form expression
\begin{multline}
\bar{n}(z,s)=\alpha e^{\lambda z}\biggl[e^{-\beta|z-z_{0}|}-e^{-\beta|z+z_{0}|}\\
  -\frac{4\sinh\left(\beta z\right)\sinh\left(\beta z_{0}\right)}{e^{2\beta L}-1}\biggr].\label{eq:SolnD}
\end{multline}
A necessary condition for convergence to the closed form expression
Eq. \eqref{eq:SolnD} is
\begin{equation}
|\exp(-2\beta L)|<1,\label{eq:nbar-closedform-condition}
\end{equation}
which defines the region of convergence of the Laplace domain function
Eq. \eqref{eq:SolnD}.

It should be emphasized that Eq. \eqref{eq:SolnD} is a general result,
valid for fractional and non-fractional cases. For normal transport
(i.e., crystalline semiconductors or gaseous electronics), $\gamma=1$,
and Eq. \eqref{eq:SolnC} has an analytic inverse Laplace transform
that reduces to Eq. (7) of \cite{Robson1985}, where it was obtained
using time domain methods. For dispersive transport, $\gamma<1$,
and an analytical inverse Laplace transform is difficult to find,
so the applications presented below required numerical inversion of
the Laplace transform%
\footnote{Numerical inverse Laplace transformation was achieved using Matlab
code published on the Mathworks File Exchange by W. Srigutomo \cite{Srigutomo2006}.
For large values of the parameter $\beta$ (defined in Eq. \eqref{eq:beta}),
the Multiple Precision (MP) Toolbox for Matlab \cite{Barrowes2009}
was required to obtain numerical convergence. The MP Toolbox uses
the open-source GNU Multiple Precision Arithmetic Library (http://gmplib.org/).%
}.

\section{Currents and the Sum Rule}

\subsection{Number, number density and charge carrier current in the time of
flight experiment }

A typical time of flight experiment measures the external current
as photogenerated carriers are driven through the sample by an applied
electric field. Under the condition that the experimental time scale
is much less than the RC time of the measurement circuit, the observed
current is the space averaged conduction current
\begin{equation}
I=\frac{1}{L}\int_{0}^{L}j(z,t)dz.
\end{equation}

Expressed in terms of the number density $n(z,t)$, the photocurrent
is 
\begin{equation}
I(t)=q\frac{d}{dt}\left\{ \frac{1}{L}\int_{0}^{L}zn(z,t)dz-\int_{0}^{L}n(z,t)dz\right\} ,\label{eq:CurrentFormula}
\end{equation}
where $q$ is the charge on each carrier. The origin of Eq. \eqref{eq:CurrentFormula}
is detailed in Appendix \ref{sec:Appendix}. Substituting the time
domain $n(z,t)$ solution Eq. \eqref{eq:SolnA} into Eq. \eqref{eq:CurrentFormula},
the current is found to be
\begin{equation}
I(t)=\sum_{m=1}^{\infty}\kappa_{m}t^{-1}E_{\gamma,0}\left(-\omega_{m}t^{\gamma}\right),\label{eq:Current}
\end{equation}
with 
\begin{multline*}
\kappa_{m}=\frac{2qn_{0}e^{-\lambda z_{0}}k_{m}D_{L}}{L^{2}\omega_{m}^{2}}\sin\left(k_{m}z_{0}\right)\\
\times\left[2\lambda D_{L}\left(e^{\lambda L}\left(-1\right)^{m}-1\right)-L\omega_{m}\right].
\end{multline*}

Alternatively, a closed form expression may be found in Laplace space
by substituting Eq. \ref{eq:SolnD} into Eq. \eqref{eq:CurrentFormula}.

\subsection{Sum rule for asymptotic slopes}

Experimental time of flight current traces plotted on double logarithmic
axes often demonstrate two distinct straight line regimes (see, for
example, Figure \ref{fig:expdata_smo}), a distinctive shape which
has been described as the {}``signature'' of dispersive transport
\cite{Scher1991}. In many materials, the sum of the slopes on logarithmic
axes of these two regimes is very close to $-2$ (Refs. \cite{Scher1991,Zallen1983}),
a prediction originally made for a continuous time random walk model
by Scher and Montroll \cite{Scher1975}. In what follows, we prove
that our expression for the current, Eq. \eqref{eq:Current}, demonstrates
the same {}``sum of slopes'' criterion.

The small argument asymptote of the Mittag-Leffler function can be
written down from its power series definition, Eq. \eqref{eq:mlf-power-series}.
The result is
\[
E_{\gamma,0}(-\omega_{m}t^{\gamma})\sim-\omega_{m}t^{\gamma},
\]
where we have neglected terms of order $O\left(\left[\omega_{m}t^{\gamma}\right]^{2}\right)$
and higher. Substituting this into Eq. \eqref{eq:Current} we find
the early time current to be
\begin{eqnarray*}
I_{\text{early}}(t) & \approx & \sum_{m=1}^{\infty}-\kappa_{m}t^{-1}\omega_{m}t^{\gamma}\sim t^{\gamma-1}.
\end{eqnarray*}
Conversely, for the long time current, we use the large $|z|$ asymptote
valid for negative real $z$ \cite{Gorenflo2002}
\[
E_{\alpha,\beta}(z)=-\sum_{k=1}^{p}\frac{z^{-k}}{\Gamma\left(\beta-\alpha k\right)}+O\left(\left|z\right|^{-1-p}\right).
\]
If $t$ is large, then by taking $p=1$ we obtain the following form
for the long time current
\begin{eqnarray*}
I_{\text{late}}(t) & \approx & -\sum_{m=1}^{\infty}\kappa_{m}t^{-1}\frac{\left(-\omega_{m}t^{\gamma}\right)^{-1}}{\Gamma\left(-\gamma\right)}\\
 & \sim & t^{-(1+\gamma)},\qquad\gamma\neq1.
\end{eqnarray*}
In summary, the asymptotic forms of the current for $\gamma\neq1$
are
\begin{equation}
I(t)\sim\begin{cases}
t^{-(1-\gamma)}, & \text{early times}\\
t^{-(1+\gamma)}, & \text{late times},
\end{cases}\label{eq:CurrentAsymptotes}
\end{equation}
in agreement with the sums of slopes condition. 

It is noteworthy that these asymptotes are \emph{independent of the
boundary conditions} imposed on the system. When solving the fractional
diffusion equation $n(z,t)$ is assumed to be factorable as $n(z,t)=Z(z)T(t)$.
The time-dependent function, $T(t)$ can be expressed in terms of
Mittag-Leffler functions 
\begin{equation}
n(z,t)=\sum_{m}Z_{m}(z)E_{\gamma}\left(c_{m}t^{\gamma}\right),\label{eq:n_without_bc}
\end{equation}
where $c_{m}$ are the separation eigenvalues found by applying the
boundary conditions to the differential equation for $Z(z)$. The
asymptotes of the Mittag-Leffler functions \cite{Gorenflo2002} are
such that physically acceptable solutions must have $c_{m}<0$ so
that $n(z,t)$ remains bounded as $t\to\infty$. Imposing only the
requirement that the boundary conditions result in a negative separation
constant, using Eq. \eqref{eq:CurrentFormula} the current must take
the form
\begin{eqnarray*}
I(t) & = & \sum_{m}\left\{ t^{-1}E_{\gamma,0}\left(c_{m}t^{\gamma}\right)\int_{0}^{L}\left(\frac{z}{L}-1\right)Z_{m}(z)dz\right\} .
\end{eqnarray*}
Using the asymptotic limits detailed above, the time dependence may
be brought outside the summation, and the the same temporal asymptotes
detailed above then follow. This result is independent of the spatial
boundary conditions and hence independent of the specific form of
$Z(z)$.

\subsection{Transit Time}

The transit time can be obtained from the expression for the total
number of charge carriers within the medium. Defining 
\[
\bar{N}(s)\equiv\int_{0}^{L}\bar{n}(z,s)dz
\]
we find in Laplace space
\begin{equation}
\bar{N}=\frac{n_{0}}{s}\left[1-e^{-\left(\lambda+\beta\right)z_{0}}-\frac{\sinh\left(\beta z_{0}\right)}{\sinh\left(\beta L\right)}e^{-\lambda z_{0}}\left(e^{\lambda L}-e^{-\beta L}\right)\right].\label{eq:Nbar}
\end{equation}
To simplify the mathematics and obtain an estimate for the transit
time, we neglect diffusion by taking the limit $D_{L}\to0$: 
\begin{eqnarray}
\bar{N}_{D_{L}=0} & = & \frac{n_{0}}{s}\left(1-\exp\left[\frac{-s^{\gamma}\left(L-z_{0}\right)}{W}\right]\right).\label{eq:Nbar_NoDiffusion}
\end{eqnarray}

In the classical case with $\gamma=1$, the above equation has the
expected inverse Laplace transform
\[
N_{D_{L}=0}^{(\text{classical})}(t)=n_{0}\left[1-H\left(t-\frac{L-z_{0}}{W}\right)\right],
\]
where $H(t)$ is the Heaviside step function.

For the dispersive case, where $\gamma<1$, Laplace inversion by complex
contour integration gives
\begin{equation}
N_{D_{L}=0}(t)=n_{0}\sum_{m=1}^{\infty}\eta_{m,\gamma}\left(\frac{L-z_{0}}{Wt^{\gamma}}\right)^{m},\label{eq:N_dispersive}
\end{equation}
where 
\[
\eta_{m,\gamma}\equiv\frac{\left(-1\right)^{m+1}\sin\left(m\pi\gamma\right)\Gamma\left(\gamma m\right)}{\pi m!}.
\]
In the special case of $\gamma=1/2$, the power series Eq. \eqref{eq:N_dispersive}
is equivalent to the closed form expression
\begin{equation}
N_{D_{L}=0}^{(\gamma=0.5)}(t)=n_{0}\,\text{erf}\left(\frac{L-z_{0}}{2W\sqrt{t}}\right),\label{eq:N_gamma0.5}
\end{equation}
where erf is the Gaussian error function. It is interesting to note
that Eq. \eqref{eq:N_gamma0.5} demonstrates great dispersion despite
it being a zero diffusion limit of the true behavior of the system.

A clear transit time cannot be precisely defined because the packet
of charge carriers becomes widely dispersed. Nevertheless, there exist
two regimes of current transport behavior, and the boundary between
these regimes defines a {}``transit time'' for the material. It
can be seen that two distinct regimes will emerge from Eq. \eqref{eq:N_dispersive},
according to the magnitude of the term in parenthesis. The transit
time, defining the transition between regimes, is therefore approximately
given by
\[
\frac{L-z_{0}}{Wt_{tr}^{\gamma}}\sim1.
\]
Solving for the transit time $t_{tr}$ 
\begin{equation}
t_{tr}\sim\left(\frac{L-z_{0}}{W}\right)^{1/\gamma}.\label{eq:transit_time}
\end{equation}
This is in agreement with the expected experimental length and field
dependence \cite{Scher1991,Scher1975,Zallen1983}.

\begin{figure}[h]
  \includegraphics[width=1\columnwidth]{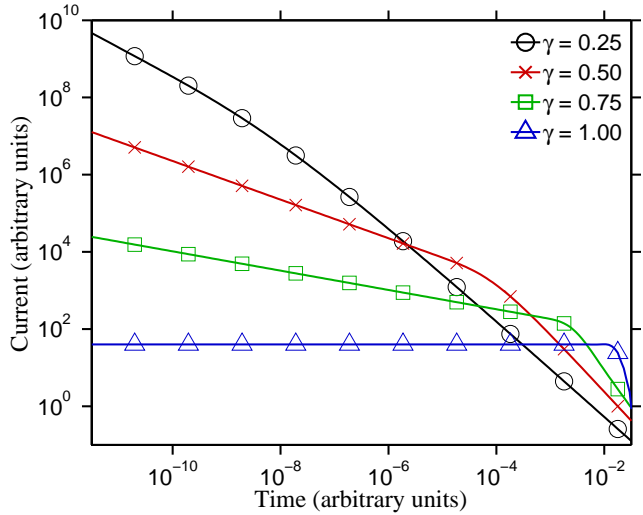}

  \caption{\label{fig:i_gamma}(Color online) Impact of the fractional order $\gamma$ on the
    temporal current profiles. Each curve is the current resulting from
    the respective number density solution of Figure \ref{fig:contour_nd_gamma}.}
\end{figure}

\begin{figure}
  \subfloat[\label{fig:contour_gamma=00003D1}$\gamma=1.00$]{\includegraphics[height=0.2\textheight]{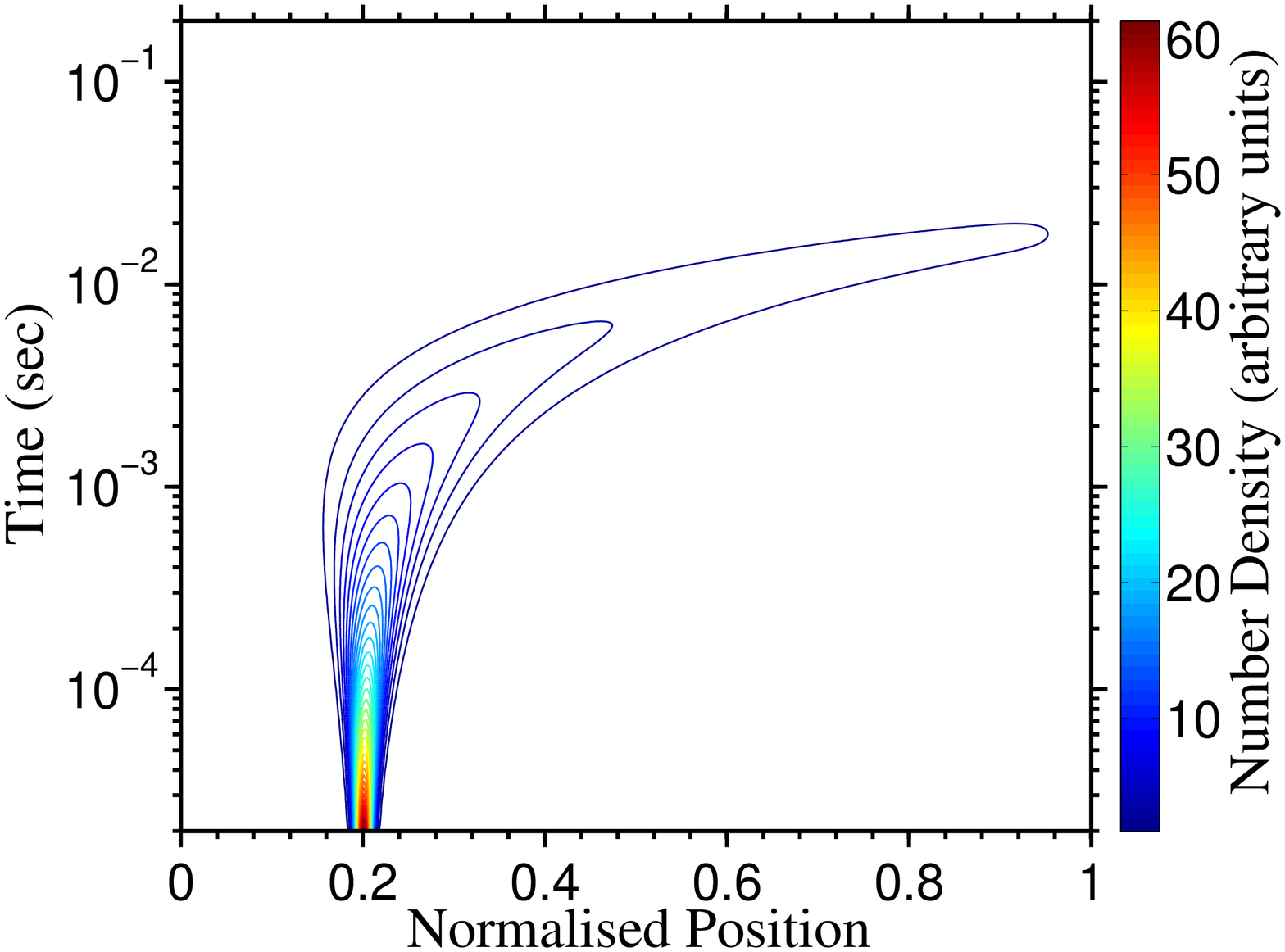}

  }

  \subfloat[\label{fig:contour_gamma=00003D0.75}$\gamma=0.75$]{\includegraphics[height=0.2\textheight]{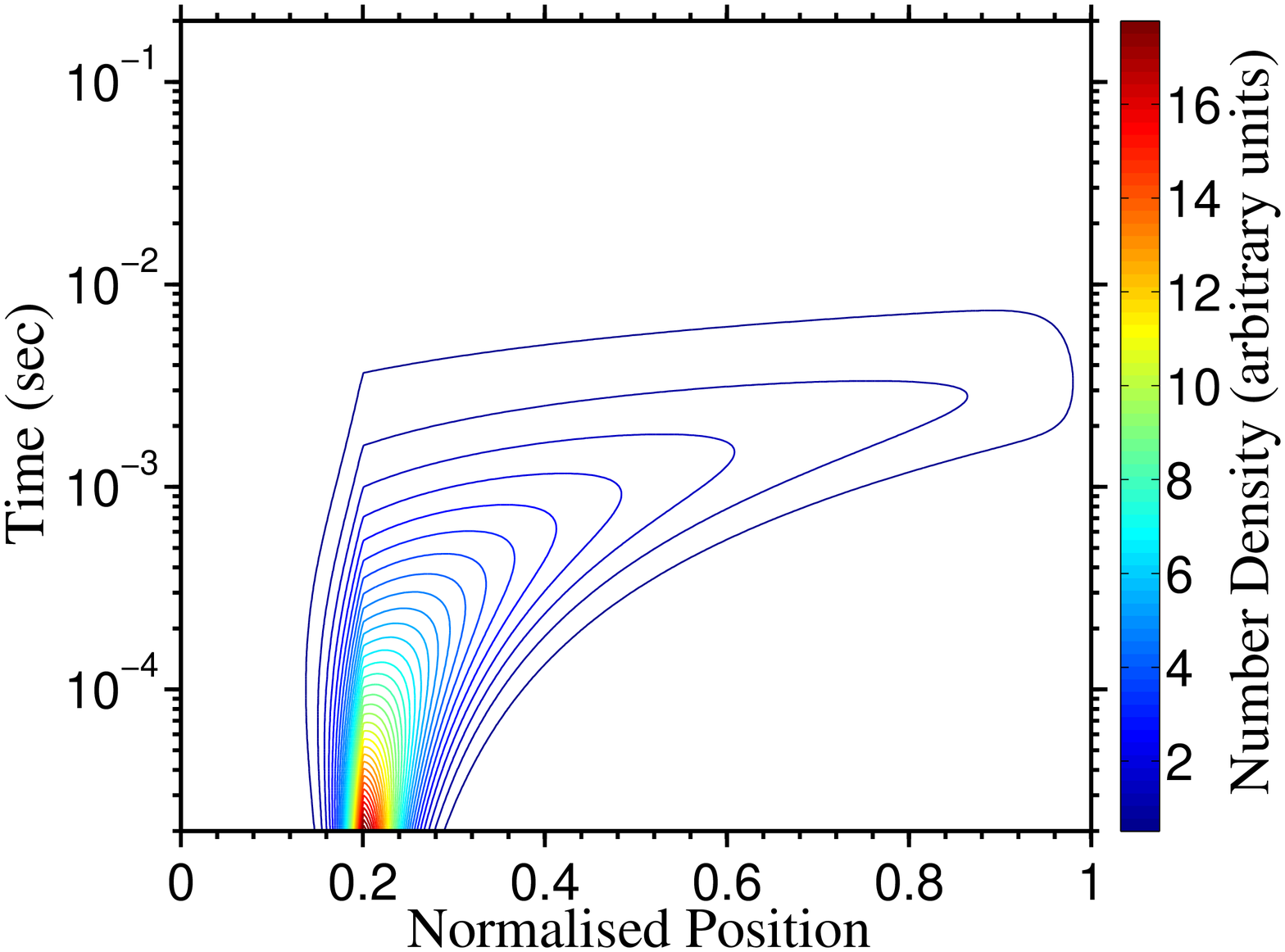}

  }

  \subfloat[\label{fig:contour_nd_gamma=00003D0.50}$\gamma=0.50$]{\includegraphics[height=0.2\textheight]{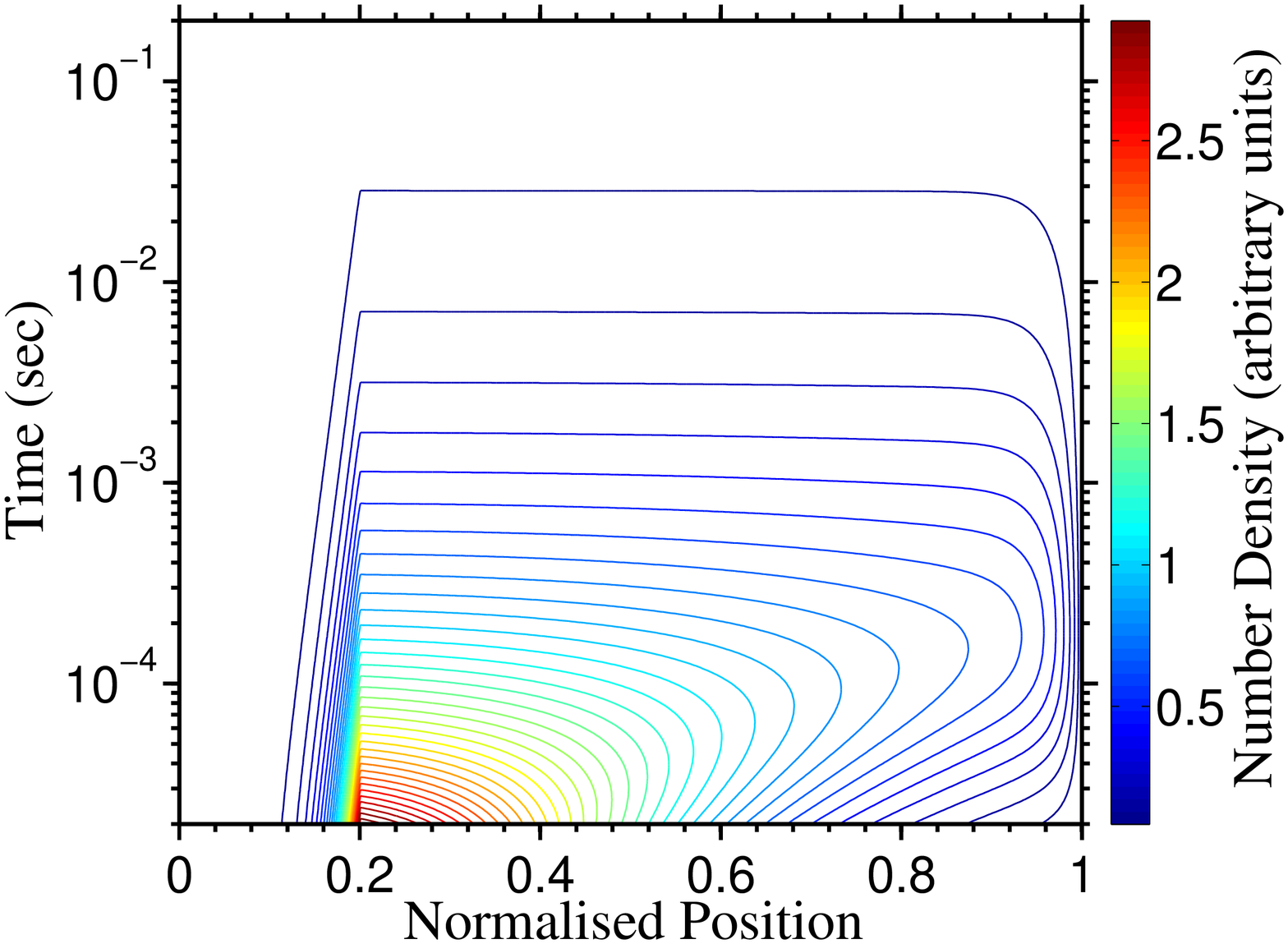}

  }

  \subfloat[\label{fig:contour_nd_gamma=00003D0.25}$\gamma=0.25$]{\includegraphics[height=0.2\textheight]{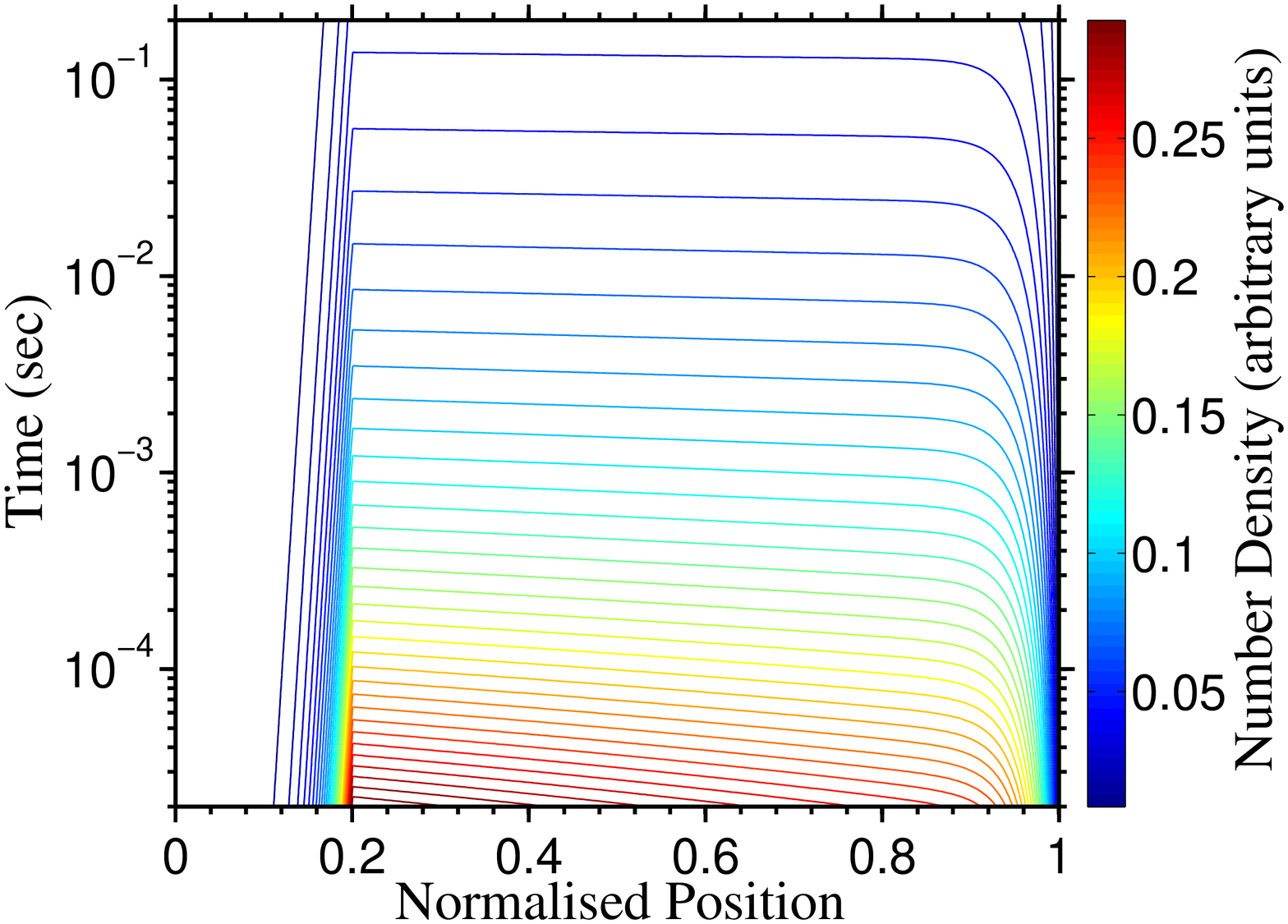}

  }

  \caption{\label{fig:contour_nd_gamma}(Color online) Impact of the fractional order $\gamma$
    of the trapping distribution on the space-time evolution of the number
    density. In these plots, $W=40/L$ ($\text{s}^{-\gamma}$) and $D_{L}=1/L^{2}$
    ($\text{s}^{-\gamma}$).}
\end{figure}

\begin{figure}[p]
  \subfloat[$W=10.0$; $D_{L}=1.0$]{\includegraphics[height=0.2\textheight]{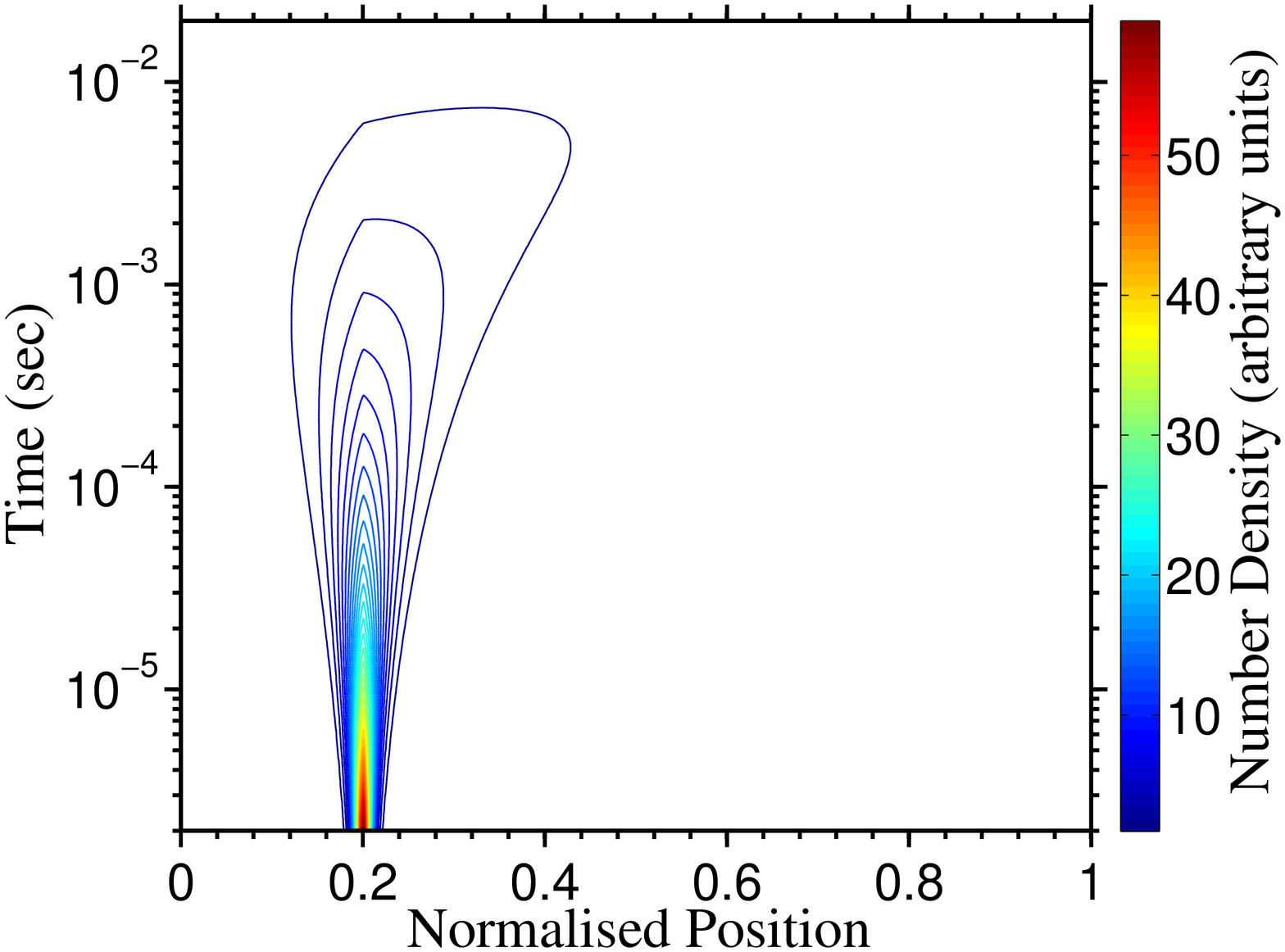}

  }

  \subfloat[$W=10.0$; $D_{L}=10.0$]{\includegraphics[height=0.2\textheight]{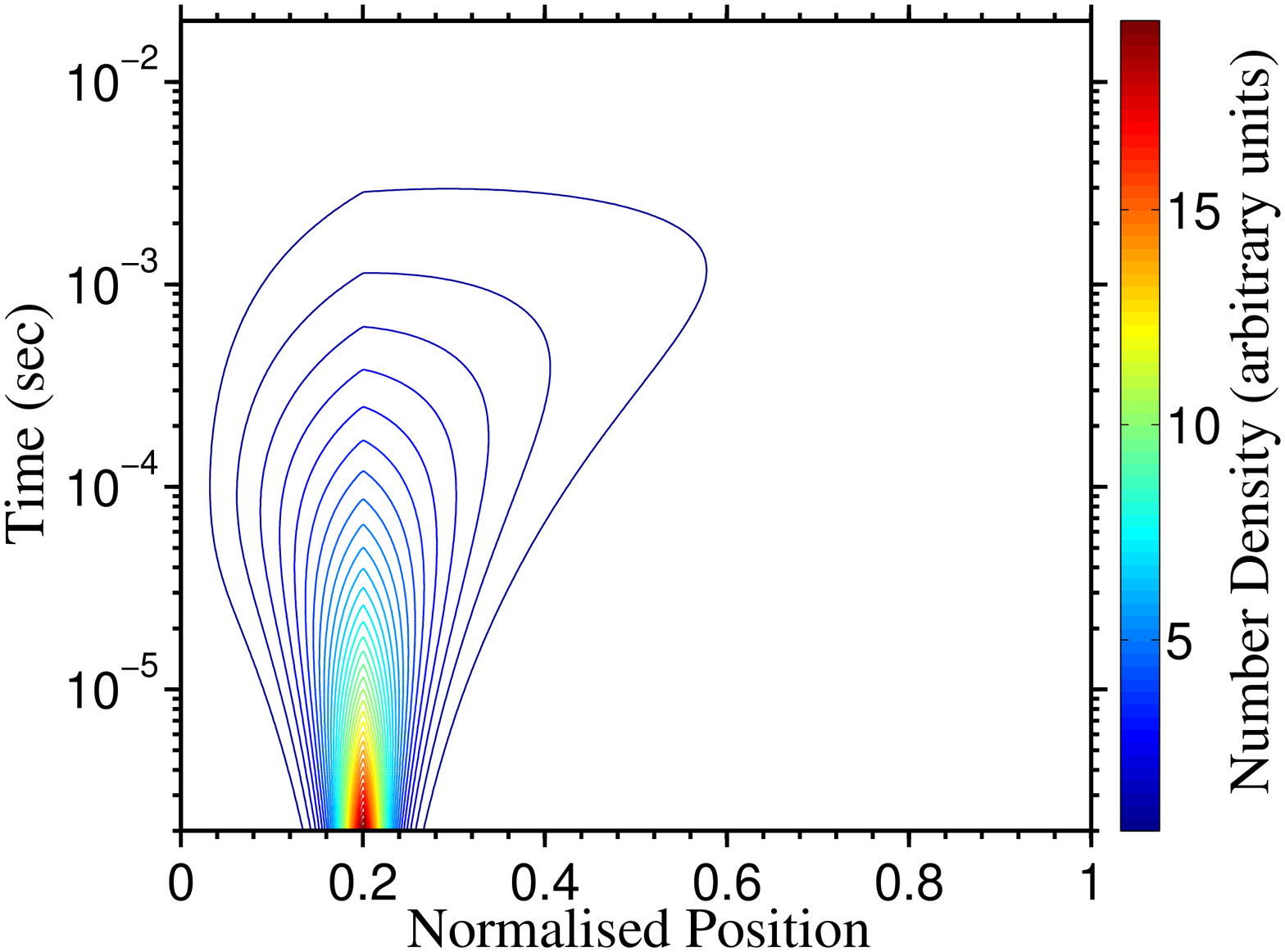}

  }

  \subfloat[$W=100.0$; $D_{L}=1.0$]{\includegraphics[height=0.2\textheight]{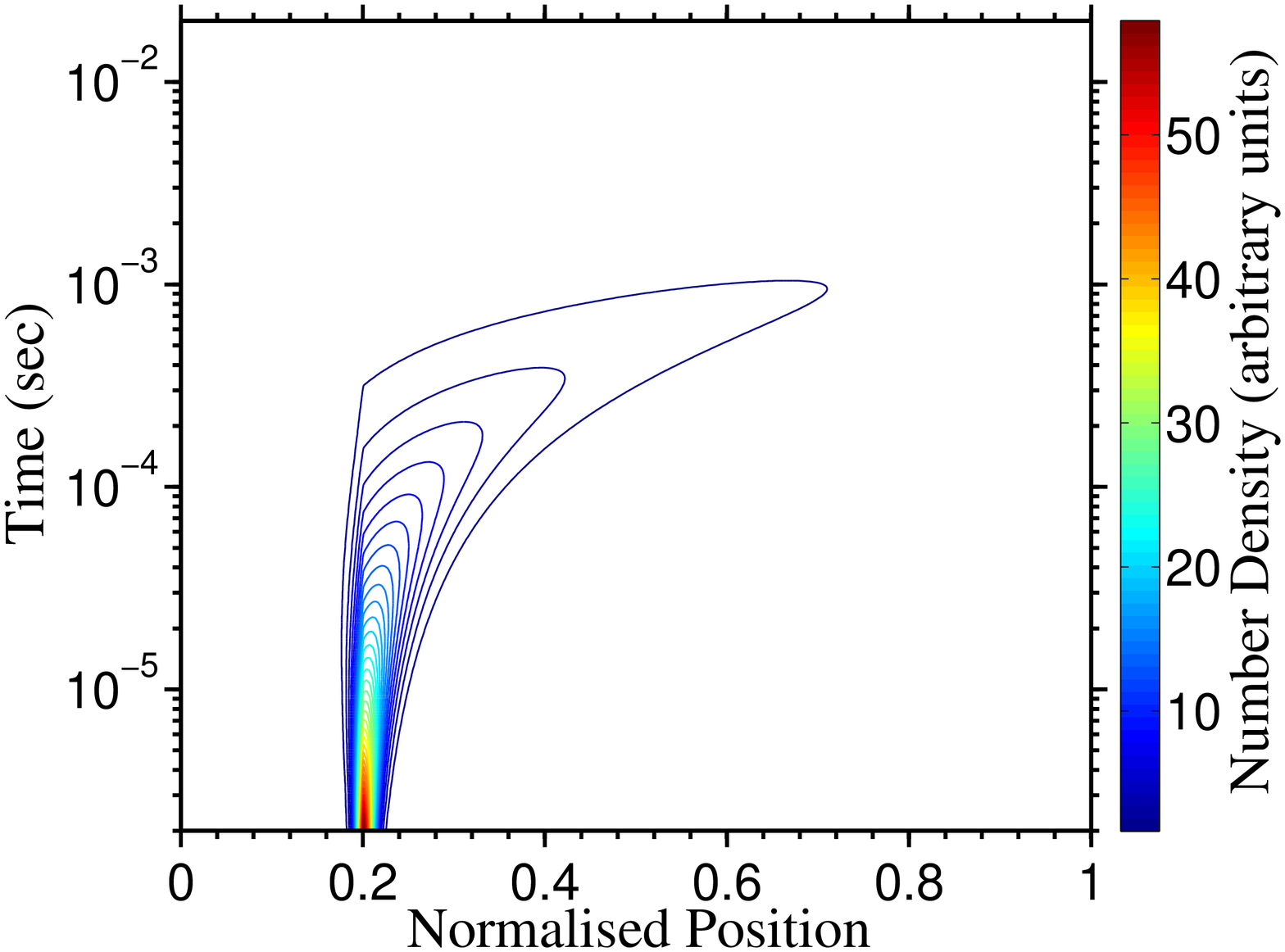}

  }

  \subfloat[$W=100.0$; $D_{L}=10.0$]{\includegraphics[height=0.2\textheight]{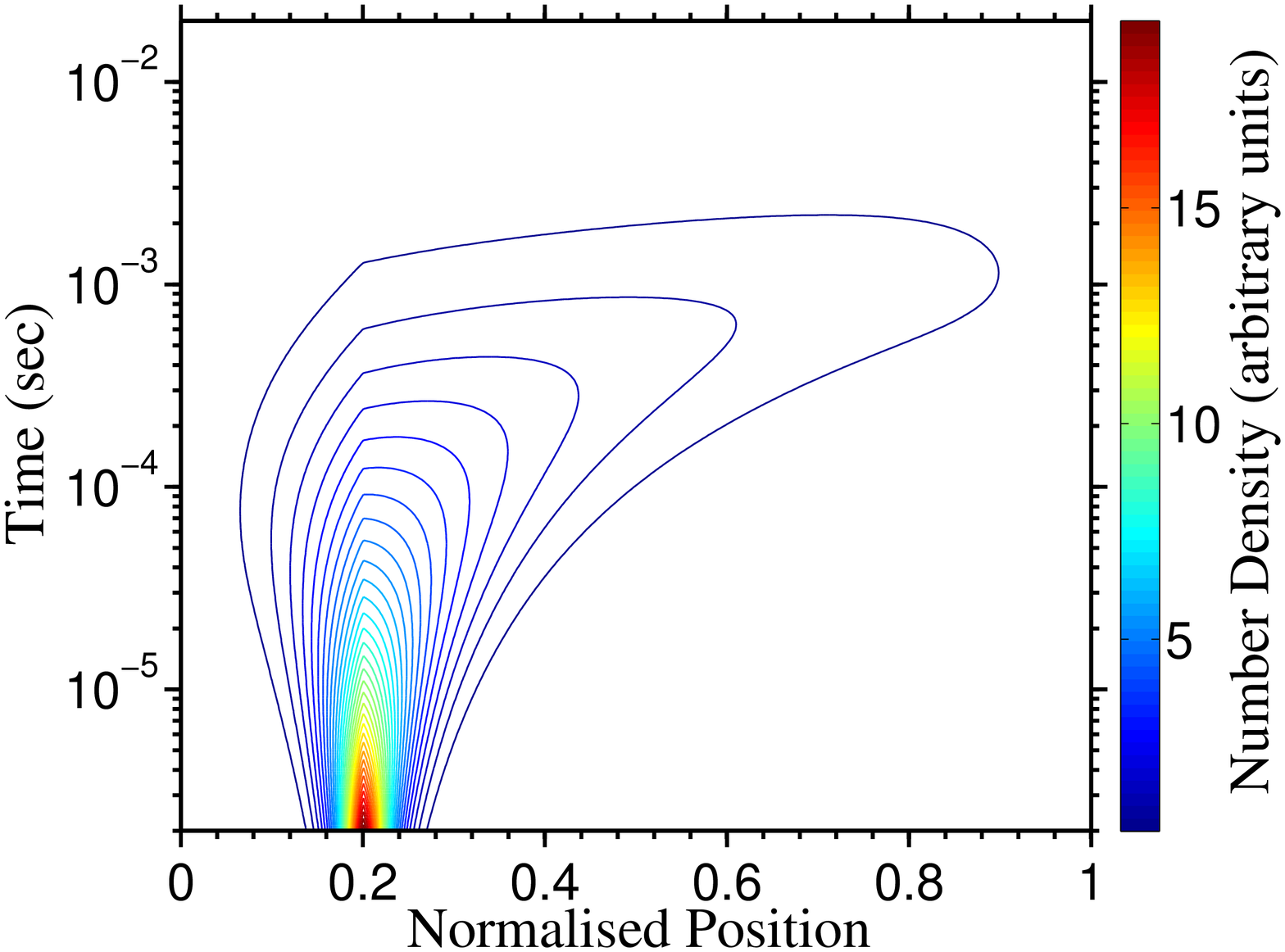}

  }

  \caption{\label{fig:contour_nd_highgamma}(Color online) Space-time evolution of the number
    density profile for $\gamma=0.8$. Here, $W$ and $D_{L}$ are normalized
    to the length of the apparatus and are hence both specified in units
    of $\text{s}^{-\gamma}$.}
\end{figure}

\begin{figure}[p]
  \subfloat[$W=10.0$; $D_{L}=1.0$]{\includegraphics[height=0.2\textheight]{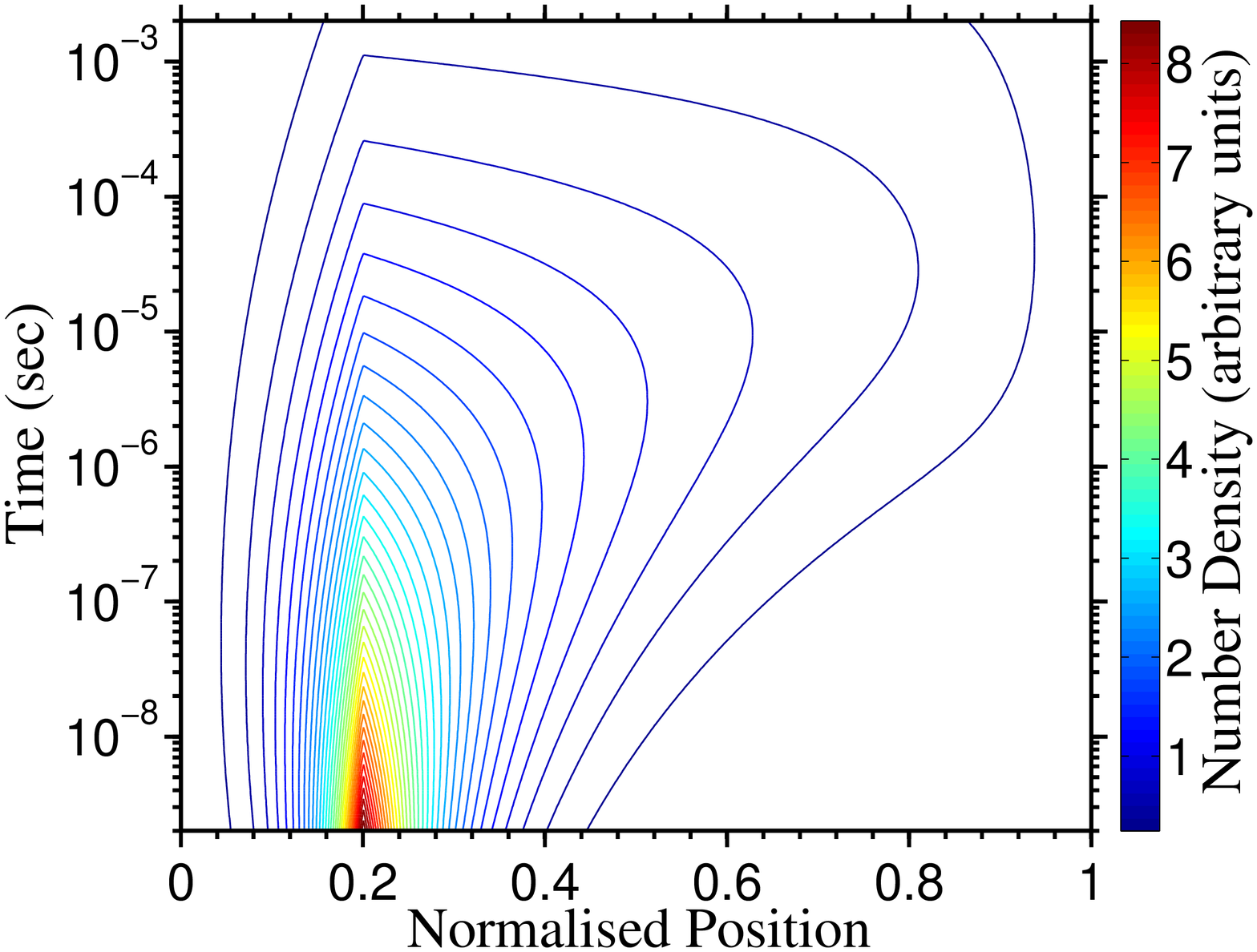}

  }

  \subfloat[$W=10.0$; $D_{L}=10.0$]{\includegraphics[height=0.2\textheight]{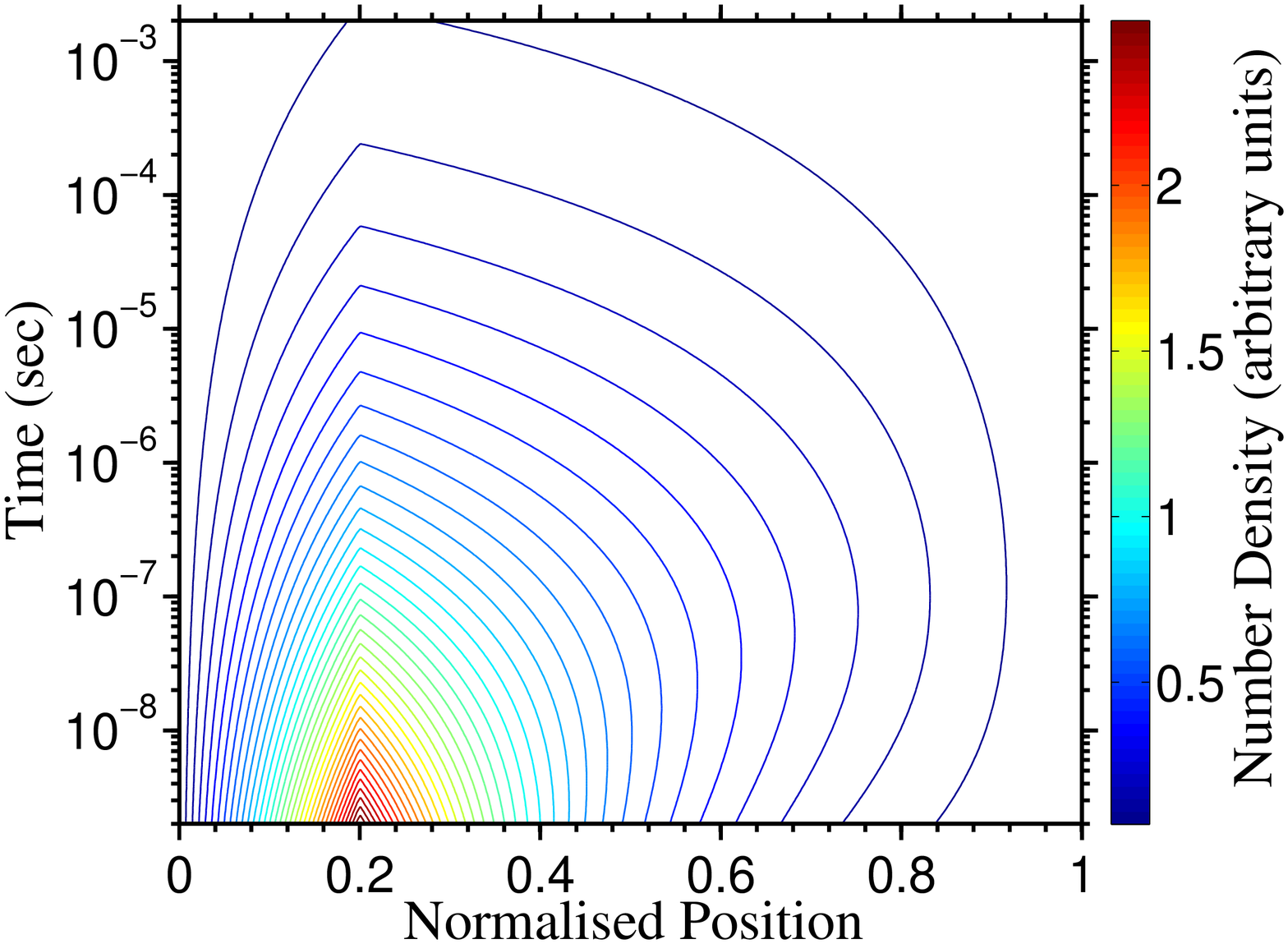}

  }

  \subfloat[$W=100.0$; $D_{L}=1.0$]{\includegraphics[height=0.2\textheight]{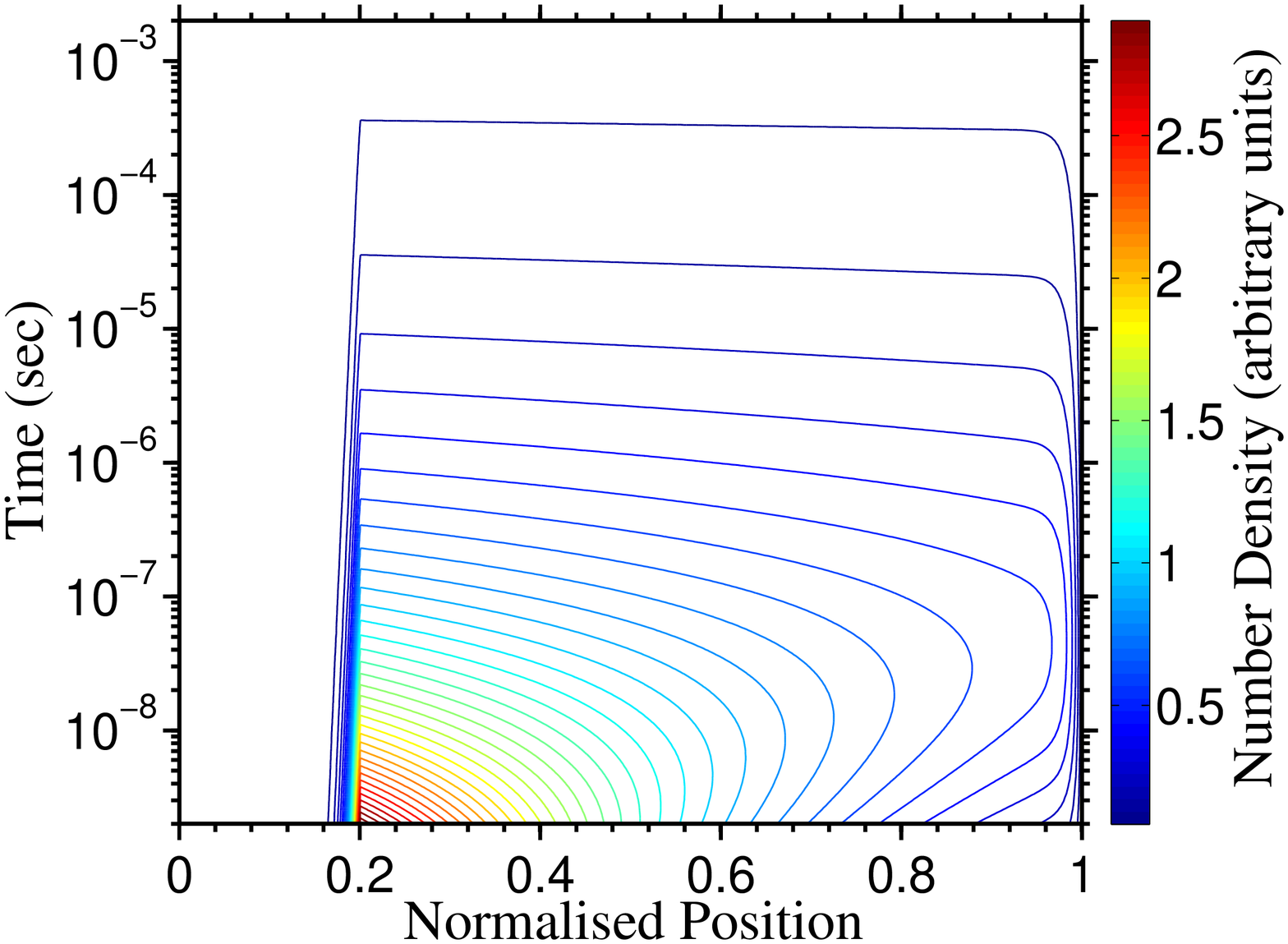}

  }

  \subfloat[$W=100.0$; $D_{L}=10.0$]{\includegraphics[height=0.2\textheight]{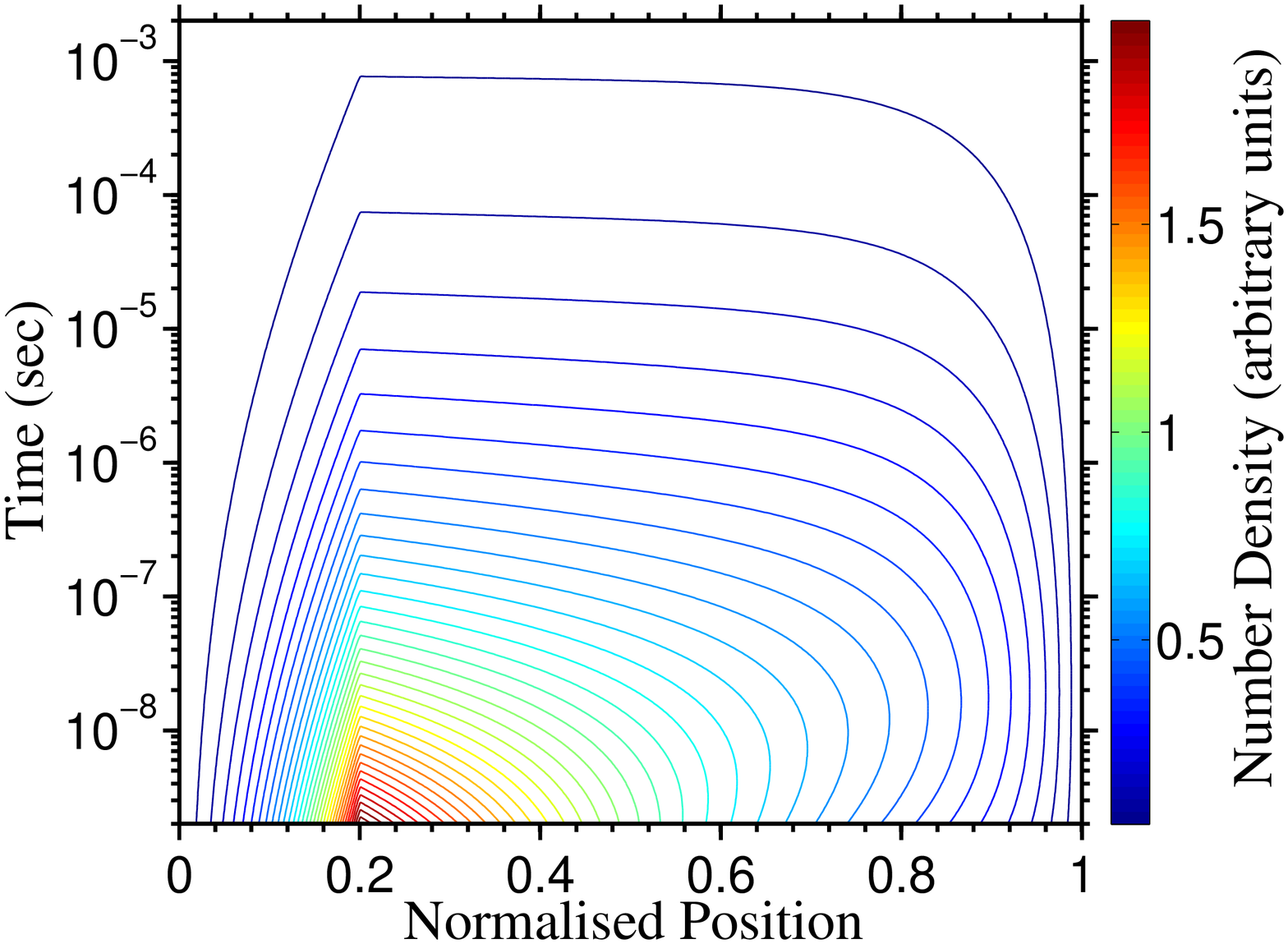}

  }

  \caption{\label{fig:contour_nd_lowgamma}(Color online) Space-time evolution of the number
    density profile for $\gamma=0.4$. Notice that these figures use a
    different time scale to those in Figure \ref{fig:contour_nd_highgamma}.
    Here, $W$ and $D_{L}$ are normalized to the length of the apparatus
    and are hence both specified in units of $\text{s}^{-\gamma}$.}
\end{figure}

\begin{figure}
  \includegraphics[width=1\columnwidth]{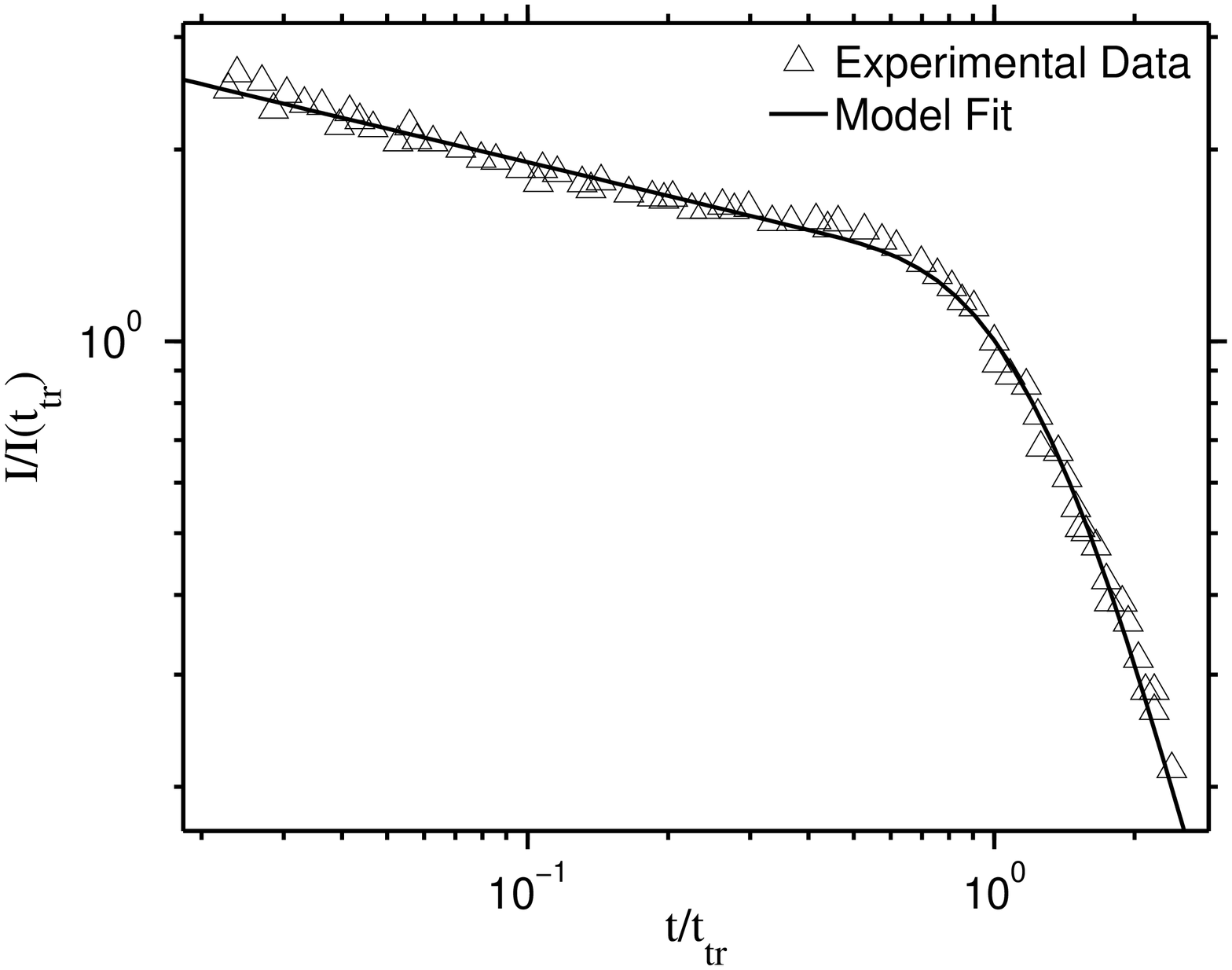}

  \caption{\label{fig:expdata_smo}Experimental time of flight current trace
    data for trinitrofluorenone-polyvinylcarbazole, digitized from \cite{Scher1975}.
    The solid line is the model fit.}
\end{figure}

\section{Results}

\subsection{Impact of model parameters on the density and current profiles}

The model discussed above has five parameters: the fractional drift
velocity $W$, the fractional diffusion coefficient $D_{L}$, the
fractional order $\gamma$, the initial source location $z_{0}$,
and the length of the sample $L$. These parameters are constrained
such that $0<\gamma\leq1$, $0<z_{0}<L$ and $D_{L}>0$. The effects
of varying the first three of these parameters will be discussed below.
The remaining two, the initial location and length of the sample,
have obvious implications for the number density profiles.

\subsubsection{Variation in fractional order $\gamma$}

The fractional order $\gamma$ is a dimensionless quantity which defines
the degree of the trapping within the medium, with a smaller value
corresponding to greater and longer lasting traps. The maximum value
of $\gamma=1$ corresponds to {}``normal transport,'' which is governed
by the classical (non-fractional) diffusion advection equation.

The impact of $\gamma$ on the electric current is demonstrated in
Figure \ref{fig:i_gamma}. For non-dispersive transport ($\gamma=1$),
the result is essentially a time independent (displacement) current
until a sharp cutoff where the charged particles exit the system through
the electrode. The finite drop off time is a reflection of the diffusion
in the system. For dispersive transport, the departure of the current
traces from the classical profiles is enhanced as the fractional order
decreases. The fractional order $\gamma$ defines the slopes of the
two regimes, and hence, characterizes the fundamental \emph{shape}
of the current trace. The relevant relations are given in Eq. \eqref{eq:CurrentAsymptotes}
above.

Number density profiles corresponding to the aforementioned current
solutions are shown in Figure \ref{fig:contour_nd_gamma}. Solutions
for $\gamma=1$ exhibit a moving Gaussian {}``pulse'' of charge
carriers, spreading according to $D_{L}$ and drifting according to
$W$. This is shown in Figure \ref{fig:contour_gamma=00003D1}.

For $\gamma<1$, the signature of fractional or dispersive behavior
appears. In this mode, the number density profile retains a {}``memory''
of the initial sharp spike at $z=z_{0}$. This peak in the density
profile does not drift with $W$, as it does in the non-dispersive
case. This long persistence of the initial condition has previously
been mentioned in the literature \cite{Metzler2000,Scher1975,Sibatov2007}.
The smaller the value of $\gamma$, the more dispersive the transport.
Indeed, for strongly dispersive systems, the spike at $z=z_{0}$ is
the most prominent feature of the entire charge distribution for much
of its lifetime. This sharp spike is most clearly illustrated in the
contour plots of Figures \ref{fig:contour_nd_gamma=00003D0.50} and
\ref{fig:contour_nd_gamma=00003D0.25}.

\subsubsection{Impact of the drift velocity $W$ and diffusion coefficient $D_{L}$}

The fractional drift velocity has units of $\text{m/\text{s}}^{\gamma}$,
and describes the tendency of the charged particles to drift in the
positive $z$ direction. The fractional diffusion coefficient has
units of $\text{m}^{2}/\text{s}^{\gamma}$, and describes the tendency
of the charged particles to diffuse down the concentration gradient.
The effects of varying $W$ and $D_{L}$ are demonstrated in Figure
\ref{fig:contour_nd_highgamma}, for a weakly dispersive system $(\gamma=0.8)$;
and in Figure \ref{fig:contour_nd_lowgamma} for a strongly dispersive
system $(\gamma=0.4)$. The relevant parameters are indicated in the
figure captions. For both systems, an increased $W$ sweeps the charge
carriers further to the right, and an increased $D_{L}$ spreads the
swarm over a wider area.

\subsection{Experimental Results}

To demonstrate the process by which this model may be fitted to time-of-flight
experimental data, we consider the data for trinitrofluorenone and
polyvinylcarbazole (TNF-PVK) presented as Figure 6 of \cite{Scher1975}.
The data was digitized from the scanned plot, and the slopes of the
two regimes was used to furnish an estimate for $\gamma$. We used
$L=1$ to give a normalized length scale; and selected the initial
source location $z_{0}$ to be $0.2$, since the model is largely
insensitive to the location of the source, provided it is sufficiently
far from the electrodes to avoid substantial {}``back diffusion.''

The intercept of the two straight lines was taken to be the transit
time $t_{tr}$, and the following equation was used to furnish an
estimate of W, which provided a starting point for curve fitting:
\begin{equation}
t_{tr}\sim\frac{1}{2}\left(\frac{L-z_{0}}{W}\right)^{\frac{1}{\gamma}},\label{eq:ttr_estimate}
\end{equation}
the factor of $1/2$ being an empirical correction that gives better
results when compared with the order of magnitude estimate Eq. \eqref{eq:transit_time}.
The final remaining parameter was initially taken as $D_{L}\approx W/20$.

The parameter estimates discussed above were used as the starting
point for nonlinear least squares curve fitting. The Matlab Curve
Fitting Toolbox was used. The result of the model fitting is shown
in Figure \ref{fig:expdata_smo}.

\section{Conclusion}

We have demonstrated a fractional advection diffusion equation modeling
the hopping transport observed in many disordered semiconductors.
We have shown that the infinite series of Fourier modes {[}Eq. \eqref{eq:SolnA}{]}
for the bounded solution can be collapsed into a closed form expression
using the Poisson summation theorem {[}Eq. \eqref{eq:SolnD}{]}. It
this closed form expression that then facilitates the extraction of
model parameters from the experimental data using a simple curve fitting
routine. We have modeled a time of flight experiment by assuming the
initial condition $n(z,t_{0})=n_{0}\delta(z-z_{0})$. We have calculated
the resultant electric current, and shown that the sum of slopes on
logarithmic axes is $-2$, as predicted by other models and as verified
by experiment. It is possible to extend this solution to sources of
finite duration or finite width, by integrating with respect to $t_{0}$
or $z_{0}$, respectively. 

\begin{acknowledgments}
 The authors would like to that the financial support of the Australian Research Council Centres of Excellence program and the Smart Futures Fund's NIRAP scheme.
 \end{acknowledgments}

\appendix

\section{Caputo and Riemann-Liouville forms of the Fractional Advection Diffusion
equation\label{sec:Caputo-and-Riemenn-Liouville-1}}

\subsection{Fractional Derivatives}

The two forms of fractional derivative commonly used to describe subdiffusive
systems are the Caputo derivative and the Riemann-Liouville derivative.
In what follows, we describe fractional \emph{partial} derivatives
with respect to $t$ in terms of an arbitrary function $f(t,x,y,...)$.
For clarify of presentation, the functional dependence of $f$ on
the other variables is suppressed, and we write simply $f(t)$.

The Caputo derivative of order $0<\alpha<1$ is defined as \cite{Gorenflo2008}:
\begin{equation}
_{0}^{C}D_{t}^{\alpha}f(t)\equiv\frac{1}{\Gamma(1-\alpha)}\int_{0}^{t}\left(t-\tau\right)^{-\alpha}f'(\tau)d\tau,
\end{equation}
where $f'(\tau)$ is the ordinary partial derivative $\partial f/\partial t$
evaluated at $t=\tau$. The Laplace transform of the Caputo derivative
is:
\begin{equation}
\int_{0}^{\infty}e^{-st}\,{}_{0}^{C}D_{t}^{\alpha}f(t)\, dt=s^{\alpha}\bar{f}(s)-s^{\alpha-1}f(0),
\end{equation}
where $\bar{f}(s)$ is the Laplace transform of $f(t)$, and $f(0)$
is the initial condition.

The Riemann-Liouville fractional derivative of order $0<\alpha<1$
is the defined as \cite{Gorenflo2008}:
\begin{equation}
_{0}^{RL}D_{t}^{\alpha}f(t)\equiv\frac{1}{\Gamma(1-\alpha)}\frac{\partial}{\partial t}\int_{0}^{t}\left(t-\tau\right)^{-\alpha}f(\tau)d\tau.
\end{equation}
The Laplace transform of a Riemann-Liouville derivative is: 
\[
\int_{0}^{\infty}e^{-st}\,{}_{0}^{RL}D_{t}^{\alpha}f(t)\, dt=s^{\alpha}\bar{f}(s)-f_{0},
\]
where $f_{0}$ is a fractional initial condition:
\begin{equation}
f_{0}\equiv\frac{1}{\Gamma\left(1-\alpha\right)}\lim_{t\to0}\int_{0}^{t}\frac{f(\tau)}{\left(t-\tau\right)^{\alpha}}d\tau.\label{eq:fract-initial-cond}
\end{equation}

\subsection{Fractional Advection-Diffusion Equations}

The first model for dispersive transport was due to Scher and Montroll
\cite{Scher1975}, who used a continuous time random walk (CTRW) where
the waiting time probability density function has divergent mean.
A continuous time random walk is characterized by a hopping probability
density function (pdf) $\psi(z,t)$. We consider the decoupled case
$\psi(z,t)=\lambda(z)w(t)$ where $\lambda(z)$ is the jump length
pdf and $w(t)$ is the waiting time pdf. Under these conditions, the
CTRW has the Fourier-Laplace space solution \cite{Klafter1987}:
\begin{equation}
\bar{n}(k,s)=\frac{1-\bar{w}(s)}{s}\frac{n_{0}(k)}{1-\lambda(k)\bar{w}(s)},\label{eq:ctrw-soln}
\end{equation}
where Fourier transformed functions are denoted by explicit dependence
on the Fourier variable $k$, and $n_{0}(k)$ is the Fourier transformed
initial condition.

We postulate a CTRW where the waiting time pdf has divergent mean.
Such a pdf has the small $s$ asymptote \cite{Klafter1987,Metzler2000}:
\begin{equation}
\bar{w}(s)\sim1-\left(\tau s\right)^{\gamma}.
\end{equation}
We further postulate a well-behaved jump length pdf with moment generating
function
\[
M_{\lambda}(\chi)=1+M_{1}\chi+\frac{M_{2}\chi^{2}}{2!}+...,
\]
for first and second moments $M_{1}$ and $M_{2}$, respectively.
This corresponds to a characteristic function (i.e. Fourier transform)
in the small $k$ limit of:
\begin{equation}
\lambda(k)=M_{\lambda}(ik)\sim1+iM_{1}k-\frac{M_{2}k^{2}}{2}.
\end{equation}

Substituting these asymptotes into Eq. \eqref{eq:ctrw-soln}, and
discarding terms of order $O(ks^{\gamma})$ and higher, we obtain:
\begin{eqnarray}
\bar{n}(k,s) & = & \frac{n_{0}(k)s^{\gamma-1}}{s^{\gamma}-iWk+D_{L}k^{2}},\label{eq:propagator-a}
\end{eqnarray}
where $W\equiv M_{1}/\tau^{\gamma}$ and $D_{L}\equiv M_{2}/2\tau^{\gamma}$.
Equation \eqref{eq:propagator-a} is the free-space propagator of
fractional advection-diffusion. By rearranging Eq. \eqref{eq:propagator-a},
one can derive various forms of fractional advection diffusion equation.
For example, one readily obtains:
\begin{equation}
s^{\gamma}\bar{n}(z,s)-s^{\gamma-1}n_{0}(z)+\left(W\frac{\partial}{\partial z}-D_{L}\frac{\partial^{2}}{\partial z^{2}}\right)\bar{n}(z,s)=0,\label{eq:caputo-laplace}
\end{equation}
which is the Laplace transform of the Caputo fractional equation \eqref{eq:caputo-fade}.
Alternatively, Eq. \eqref{eq:propagator-a} may be rearranged to give:
\begin{equation}
\bar{n}(z,s)-\frac{n_{0}(z)}{s}+s^{-\gamma}\left(W\frac{\partial}{\partial z}-D_{L}\frac{\partial^{2}}{\partial z^{2}}\right)\bar{n}(z,s)=0,\label{eq:metzler-laplace}
\end{equation}
which is a fractional integral equation. Inverting the Laplace transform in
Eq. \eqref{eq:metzler-laplace}, and taking an ordinary partial derivative with respect to
time, one obtains the following form of the fractional advection diffusion equation:
\begin{equation}
\frac{\partial n}{\partial t}+\,_{0}^{RL}D_{t}^{1-\gamma}\left(W\frac{\partial n}{\partial z}-D_{L}\frac{\partial^{2}n}{\partial z^{2}}\right)=0.\label{eq:metzler-fade}
\end{equation}

Equation \eqref{eq:metzler-fade} is a special case of the fractional
Fokker-Planck equation \cite{Metzler1999,Barkai2001}, and is equivalent
to the Caputo fractional advection diffusion equation \eqref{eq:caputo-fade}
considered in this paper.

\section{Derivation of Current Formula\label{sec:Appendix}}

\begin{figure}
  \includegraphics[width=1\columnwidth]{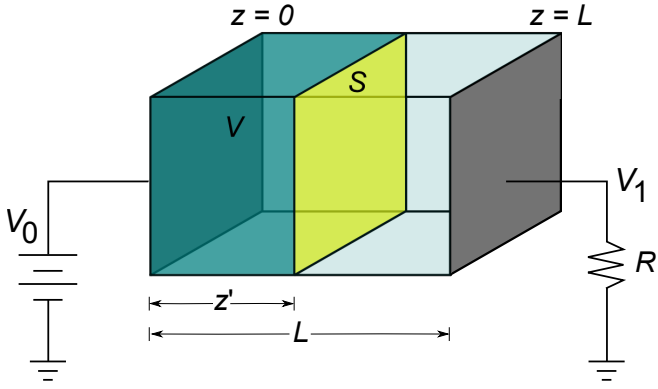}

  \caption{\label{fig:ToF-System}(Color online) Simplified time of flight schematic used in
    current derivation. The two electrodes at $z=0$ and $z=L$ have potentials
    $V_{0}$ and $V_{1}$, respectively. A surface $\mathcal{S}$ cuts
    through the sample at $z=z'$; the volume $\mathcal{V}$ is the space
    between the $z=0$ electrode and the surface $\mathcal{S}$.}
\end{figure}

Consider a time of flight system where all spatial variation is confined
to the $z$ direction, normal to the electrodes. An electrode at $z=0$
is held at a potential $V_{0}$ by an external power supply, and the
opposite electrode at $z=L$ has potential $V_{1}$ and is connected
via a resistor $R$ to the ground, as shown in Figure \ref{fig:ToF-System}.
We define a surface $\mathcal{S}$ which is normal to the electrodes
at a position $z=z'$, and a volume $\mathcal{V}$ which is the entire
area between the $z=0$ electrode and the surface $\mathcal{S}$.

The overall current will consist of a conduction current and a displacement
current. Integrating across the width of the sample:

\begin{equation}
I=\frac{1}{L}\int_{0}^{L}j(z',t)dz'+\frac{\epsilon A}{L}\frac{d}{dt}\left(V_{0}-V_{1}\right),
\end{equation}
where $j(z',t)$ is the conduction current passing through the surface
$\mathcal{S}$, $\epsilon$ is the permittivity of the semiconducting
material, and $A$ is the area of the electrodes.

Under typical measuring conditions, the transit time $t_{tr}$ is
much less than the RC time of the circuit. Therefore, we assume that
$V_{0}-V_{1}$ is essentially constant, and then the current is simply
the space-averaged conduction current:
\begin{equation}
I=\frac{1}{L}\int_{0}^{L}j(z',t)dz'.\label{eq:curr_avg_j}
\end{equation}

The conduction current leaving the volume $\mathcal{V}$ is the negative
rate of change of the charge enclosed:
\[
j(z',t)=-\frac{d}{dt}\int_{0}^{z'}qn(z,t)dz.
\]
Using Eq. \eqref{eq:curr_avg_j}:
\begin{eqnarray*}
I & = & -\frac{q}{L}\frac{d}{dt}\int_{0}^{L}\int_{0}^{z'}n(z,t)dzdz'.
\end{eqnarray*}
Changing the order of integration:
\begin{eqnarray}
I & = & -\frac{q}{L}\frac{d}{dt}\int_{0}^{L}\int_{z}^{L}n(z,t)dz'dz\nonumber \\
 & = & -\frac{q}{L}\frac{d}{dt}\int_{0}^{L}\left(L-z\right)n(z,t)dz\nonumber \\
 & = & q\frac{d}{dt}\left\{ \frac{1}{L}\int_{0}^{L}zndz-\int_{0}^{L}ndz\right\} .\label{eq:appendix_current}
\end{eqnarray}

It should be noted that different expressions exist within the literature
for the current depending on whether the paper in question uses a
multiple trapping model or a hopping model. This is why our current
expression \eqref{eq:appendix_current} is at first glance not equivalent
to the current expressions used by some other authors. Under a multiple
trapping model, the equivalent is: 
\begin{equation}
I(t)\propto\frac{W}{L}\int_{0}^{L}n_{\text{free}}(z,t)dz,
\end{equation}
where $n_{\text{free}}$ is the distribution of untrapped particles
and $W$ is the drift velocity of these particles. This formula can
be obtained by neglecting diffusive flux to substitute $j=Wn_{\text{free}}$
into Eq. \eqref{eq:curr_avg_j}.

\bibliographystyle{apsrev}
%\bibliography{references}
%% bbl file pasted here

%% end of bbl file

\end{document}